\documentclass[prc,aps,floatfix,twocolumn,groupedaddress,nofootinbib,preprintnumbers,
amsmath,amssymb,amsfonts,widetable] {revtex4}
\usepackage{bm}
\usepackage{soul}
\usepackage{array}
\usepackage{lipsum}
\usepackage{relsize}
\usepackage{mathrsfs}
\usepackage{amssymb}
\usepackage{amsmath}
\usepackage{graphicx}
\usepackage[usenames]{color}
\usepackage{pslatex}
\usepackage{booktabs}
\usepackage{physics}

\begin{document}
\title{A universal reduced basis for the calibration of 
       covariant energy density functionals}
\author{Amy L. Anderson}\email{aanderson6@fsu.edu}
\author{J. Piekarewicz}\email{jpiekarewicz@fsu.edu}
\affiliation{Department of Physics, Florida State University,
               Tallahassee, FL 32306, USA}
\date{\today}
\begin{abstract}
  The reduced basis method is used to construct a ``universal" basis of Dirac orbitals
  that may be applicable throughout the nuclear chart to calibrate covariant energy 
  density functionals. Relative to our earlier work using the non-relativistic Schr\"odinger 
  equation, the Dirac equation adds an extra layer of complexity due to the existence 
  of negative energy states. However, once this problem is mitigated, the resulting 
  reduced basis is able to accurately and efficiently reproduce the high-fidelity model 
  at a fraction of the computational cost. We are confident that the resulting reduced 
  basis will serve as a foundational element in developing rapid and accurate emulators. 
  In turn, these emulators will play a critical role in the Bayesian optimization of covariant 
  energy density functionals.
\end{abstract}
\smallskip

\maketitle

\section{Introduction}
\label{Sec:Introduction}

Since first suggested by the editors of the Physical Review in 2011\,\cite{PhysRevA.83.040001}, 
the theoretical nuclear physics community has embraced the notion of including uncertainty 
estimates in papers involving theoretical calculations of physical observables. To properly 
quantify theoretical uncertainties and explore correlations among physical observables, some 
of the initial efforts were devoted to the accurate calibration of a variety of nuclear 
models\,\cite{Reinhard:2010wz, Kortelainen:2010hv,Fattoyev:2011ns,Reinhard:2013fpa,
Dobaczewski:2014jga,Piekarewicz:2014kza,Chen:2014sca,Wesolowski:2015fqa,Yuksel:2019dnp,
Drischler:2020hwi,Drischler:2020yad, Furnstahl:2020abp}. Further, some earlier applications 
of artificial neural networks used to predict nuclear masses and charge radii that lie beyond 
the experimental reach\,\cite{Athanassopoulos:2003qe,Akkoyun:2012yf,Bayram:2013hi,
Gernoth:1993}, have now been extended to include Bayesian statistics for the 
proper quantification of theoretical uncertainties\,\cite{Utama:2015hva,Utama:2016rad,
Utama:2017ytc, Neufcourt:2019qvd,Neufcourt:2020nme,Lovell:2022pkw,Saito:2023seh}. 
Finally, within the last few years, the Reduced Basis Method (RMB) have been incorporated into 
the arsenal of machine-learning tools dedicated to accelerate calculations of complex nuclear 
systems\,\cite{Frame:2017fah,Konig:2019adq,Drischler:2021qoy,Bonilla:2022rph,
Melendez:2022kid,Giuliani:2022yna,Anderson:2022jhq,Odell2024,drnuclear}. 

Perhaps the main virtue of the RBM is the ability to create a powerful framework for building efficient and 
accurate emulators. Even with advances in computer power, algorithmic developments, and physical 
insights, a full Bayesian calibration of nuclear models is generally unattainable without the help of emulators.
Besides providing robust statistical uncertainties, Bayesian inference is often used 
to identify strong correlations between a property that is inaccessible and a surrogate observable that 
may be determined experimentally. However, Bayesian inferences that rely on Monte Carlo methods 
require a large number of evaluations of the same set of observables for many different realizations of the 
model parameters. This can be computationally expensive and highly impractical. Instead, emulators 
transform such a computationally-intractable problem into a low-dimensional calculation, thereby allowing 
enormous gains in computational speed with a minimal loss in precision. In essence, the reduced basis 
method, which falls under the general rubric of ``Reduced Order Models" \cite{Brunton_Kutz_2019},  
encapsulate a set of dimensionality reduction approaches that aim at speeding up computations by 
approximating the solution of the high-fidelity problem by retaining only a handful of the most important 
components, namely, the coefficients of the reduced basis. The core assumption behind the RBM is that the 
solutions of the problem of interest vary smoothly across a subspace defined by the model parameters. 
Thus, the solution for an arbitrary set of parameters can be accurately approximated by a linear combination 
of exact solutions obtained for a small set of parameters. 

Unlike other approaches, the RBM does not rely on uncontrolled extrapolations that aim to predict the behavior 
of the system outside its range of validity. Rather, the reduced basis provides a hierarchical scheme to select 
the most important basis states. As such, the most useful part of the RBM is the ``R'', that instead of using the
entire Hilbert space, selects an optimal basis for describing the eigenstates of a Hamiltonian with an arbitrary
set of model parameters with a much smaller (reduced) set of functions. 
While the RBM have only recently been embraced by the nuclear physics community\,\cite{Bonilla:2022rph,
Melendez:2022kid}, it is a mature field that facilitates the solution of challenging numerical problems in a range 
of different areas\,\cite{Benner:2020}.

In this paper we will use eigenvector continuation\,\cite{Frame:2017fah}, a particular implementation of  the
RBM\,\cite{Quarteroni:2015,Heasthaven:2016}, that has been shown to be highly successful for calculating 
the eigenvalues and eigenvectors of a Hamiltonian matrix---even when the ``training'' set lies far away from the 
physically relevant model parameters\,\cite{Frame:2017fah}. While a direct matrix diagonalization in the entire 
Hilbert space is computationally expensive, the RBM can generate the same low-energy states with minimal 
computational burden. Several areas in nuclear physics have benefited from the development of efficient and 
accurate emulators using the RBM\,\cite{Konig:2019adq,Furnstahl:2020abp,Drischler:2021qoy}. Indeed, in a 
recent work we have constructed a ``universal'' reduced basis for the solution of a scaled Schr\"odinger equation 
that successfully reproduced the single-particle spectrum of a variety of nuclei\,\cite{Anderson:2022jhq}. Ultimately, 
however, we aim to create robust emulators for the calibration of covariant energy density functionals\,\cite{Giuliani:2022yna}. 
Hence, the extension to the relativistic domain now demands the solution of the Dirac equation. Creating a ``universal'' 
basis of Dirac orbitals to compute the single-particle structure of a variety of nuclei is the main goal of this work. 

The main justification behind this goal is twofold. First, we use the independent particle model, one of the 
fundamental pillars of nuclear structure. As argued by Bohr and Mottelson: ``the relatively long 
mean free path of the nucleons implies that the interactions primarily contribute a smoothly varying 
average potential in which the particles move independently''\,\cite{BohrI:1998}. While other 
important effects, such as pairing correlations, require a modification to the independent particle 
paradigm, we avoid these complications by focusing exclusively on the bulk properties of magic 
and semi-magic nuclei. Indeed, this kind of nuclei have been the bedrock of our calibration protocol; 
see for example\,\cite{Todd-Rutel:2005fa,Fattoyev:2010mx,Chen:2014sca,Chen:2014mza}. 
Second, Density Functional Theory (DFT) guarantees that functional minimization of a
``suitable'' energy density functional yields both the exact ground-state energy and one-body density 
of the complicated many-body system\,\cite{Hohenberg:1964zz}. However, since DFT offers no
guidance on how to construct such a suitable energy density functional, Kohn and Sham replaced
the complex interacting system by an equivalent system of \emph{non-interacting} fermions moving 
in an average potential rich enough to capture all the essential physics\,\cite{Kohn:1965}. In the next 
sections we demonstrate how a physically-motivated average potential may be used to generate a
spherical basis of Dirac orbitals that could be used through the entire nuclear chart.

The paper has been organized as follows. In Sec.\ref{Sec:Formalism} we review the structure of the 
Dirac equation and develop the formalism necessary to generate a reduced basis of Dirac orbitals.
We then test in Sec.\ref{Sec:Results} the performance of the RBM in reproducing the high-fidelity 
results. Finally, we conclude in Sec.\ref{Sec:Conclusions} with a short summary and a vision for future 
work.

\section{Formalism}
\label{Sec:Formalism}

In 1928, Paul Dirac proposed an alternative to the Schr\"odinger equation for the description
of electrons by incorporating the principles of special relativity. One of the most remarkable 
consequences of demanding that time and space be treated on equal footing is the 
unavoidable emergence of negative energy states\,\cite{Dirac:1982}. As we will show below,
the appearance of negative energy states requires special attention when constructing a 
robust reduced basis. 

\subsection{Dirac equation for spherically symmetric potentials}

As mentioned above, we aim to build a RBM that serves as input for Bayesian optimization
of covariant energy density functionals informed by the ground state properties of magic and
semi-magic nuclei. As such, the training points for building the reduced basis only require the
solution of the Dirac equation for spherically symmetric potentials with a Hamiltonian of the 
following form:
\begin{equation}
 H = {\bm\alpha}\cdot{\bf P}+ \beta\Big(M - S(r)\Big) + V(r),
 \label{Dirac}
 \end{equation}
where ${\bm \alpha}$ and $\beta$ are Dirac matrices, ${\bf P}\!=\!-i{\bm\nabla}$ is the momentum 
operator, and $M$ is the bare nucleon mass. In turn, $S(r)$ and $V(r)$ are spherically symmetric 
Lorentz scalar and time-like vector potentials that underpin the successful relativistic mean field 
theory (RMF). The hallmark of the RMF framework is strong scalar and time-like vector potentials 
that cancel to produce a relatively weak central potential but which add coherently to account for 
the strong spin-orbit splitting observed in atomic nuclei\,\cite{Horowitz:1981xw,Serot:1984ey}.

The spherical symmetry of the Hamiltonian guarantees that the eigenstates of the Dirac 
Hamiltonian can be classified according to their total angular momentum $j$ and its projection 
$m$\,\cite{Sakurai:1967}. That is,
\begin{equation}
 {\cal U}_{n \kappa m}({\bf r}) = \frac {1}{r}
 \left( \begin{array}{c}
   \phantom{i}
   g_{n \kappa}(r) {\cal Y}_{+\kappa \mbox{} m}(\hat{\bf r})  \\
  if_{n \kappa}(r) {\cal Y}_{-\kappa \mbox{} m}(\hat{\bf r})
 \end{array} \right),
\label{Uspinor}
\end{equation}
where $n$ is the principal quantum number and the spin spherical harmonics 
${\cal Y}_{\kappa \mbox{} m}$ are obtained by coupling the orbital angular $l$ 
momentum to the intrinsic nucleon spin-1/2 to obtain a total angular momentum $j$. 
That is,
\begin{equation}
 |\kappa\,m\rangle = \Big|l\frac{1}{2}jm\Big\rangle.
\end{equation}
In particular, the orbital angular momentum $l$ and the total angular momentum 
$j$ are obtained from the generalized angular momentum $\kappa$ as follows:
\begin{equation}
 j = |\kappa|\!-\!\frac {1}{2} \;; \quad
 l = \begin{cases}
               \kappa\;,  & {\rm if} \; \kappa>0; \\
        -(1+\kappa)\;, & {\rm if} \; \kappa<0.
     \end{cases}
\end{equation}
Unlike the case of the spherically symmetric Schr\"odinger equation, the orbital angular 
momentum is not a good quantum number. Instead, the orbital angular momentum of 
the upper and lower components differ by one unit.

By exploiting the spherical symmetry of the problem, one can eliminate the entire
angular dependence and reduce the problem to a set of first order, coupled differential 
equations:
\begin{subequations}
\begin{eqnarray}
  && \hspace{-25pt}
  \left(\frac{d}{dr}+\frac{\kappa}{r}\right)g_{n\kappa}(r) 
 -\left[E+M-S(r)-V(r) \right]f_{n\kappa}(r)=0,
  \label{Eq: Dirac g} \\
  && \hspace{-25pt}
  \left(\frac{d}{dr}-\frac{\kappa}{r}\right)f_{n\kappa}(r)
 +\left[E-M+S(r)-V(r)\right]g_{n\kappa}(r)=0.
  \label{Eq: Dirac f}
\end{eqnarray}
\label{DiracEqn}
\end{subequations}

In search of a universal basis, it is convenient to rescale the above equations by an 
intrinsic distance scale $c\!\propto A^{1/3}$ characteristic of each individual nucleus. 
Hence, by defining the dimensionless distance as $x\!\equiv\!r/c$, the coupled Dirac 
equation may now be written as
\begin{subequations}
\begin{eqnarray}
  && \hspace{-25pt}
  \left(\frac{d}{dx}+\frac{\kappa}{x}\right)g_{n\kappa}(x) 
 -\left[\epsilon+\mu-\sigma(x)-\omega(x) \right]f_{n\kappa}(x)=0,
  \label{DiracG} \\
  && \hspace{-25pt}
  \left(\frac{d}{dx}-\frac{\kappa}{x}\right)f_{n\kappa}(x)
 +\left[\epsilon-\mu+\sigma(x)-\omega(x)\right]g_{n\kappa}(x)=0,
  \label{DiracF}
\end{eqnarray}
\label{DiracEqns}
\end{subequations}
where we have defined,
\begin{subequations}
\begin{eqnarray}
  & \mu=cM,\hspace{5pt} & \sigma(x)=cS(r\!=\!cx), \\
  & \epsilon=cE,\hspace{5pt} & \omega(x)=cV(r\!=\!cx).
\end{eqnarray}
\label{Scaling}
\end{subequations}
Finally, we note that the Dirac orbitals satisfy the following normalization condition:
\begin{equation}
    \int_{0}^{\infty} \Big(g_{n\kappa}^2(x) +f_{n\kappa}^2(x) \Big) dx=1.
 \label{Eq: norm}
\end{equation}

\subsection{Reduced Order Models}

The main goal of this section is to illustrate how the use of a reduced order model helps 
generate efficient and accurate emulators for the solution of the Dirac equation. As 
mentioned in the Introduction, our aim is to speed up computations by approximating the 
solution of the high-fidelity problem by retaining only a handful of the most important 
components of the entire basis. But how does one identifies the most important components?

\begin{figure}[ht]
 \includegraphics[width=0.35\textwidth]{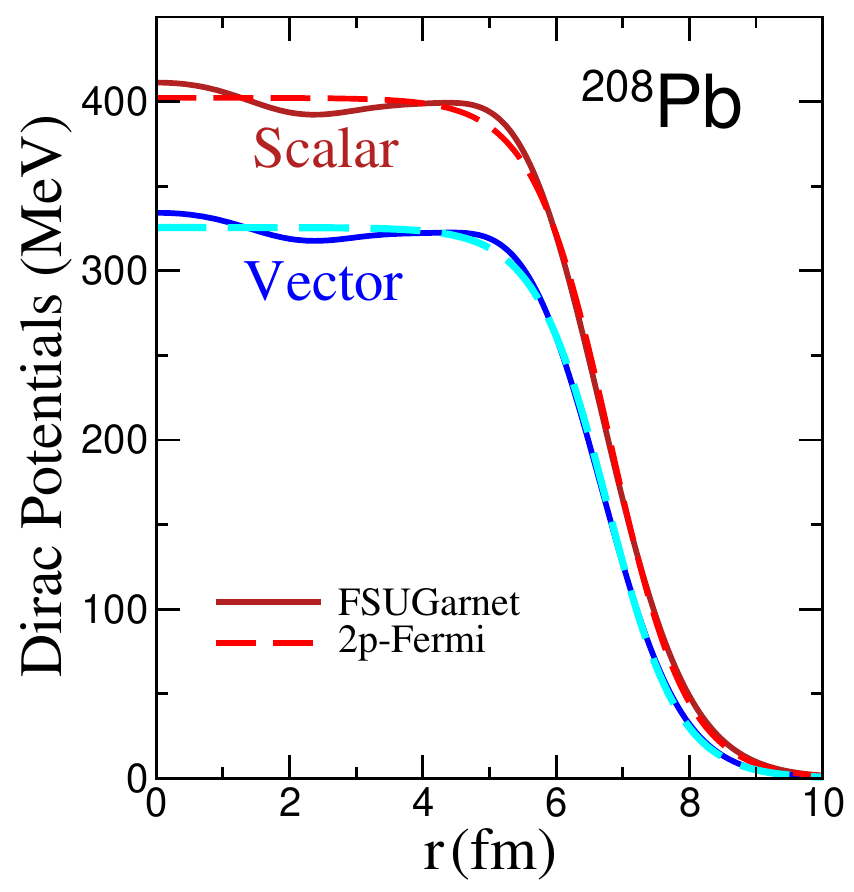}
 \caption{Self-consistently generated scalar and time-like vector potentials for $^{208}$Pb, as predicted 
               by the covariant energy density functional FSUGarnet\,\cite{Chen:2014mza}. The dashed lines 
               represent the corresponding 2-parameter Fermi fits that, after properly scaled, will be used to 
               generate the reduced basis.}
\label{Figure1}
\end{figure}

Our particular implementation of the reduced order model starts by obtaining exact solutions of 
the high-fidelity problem for a representative set of spherical nuclei. To facilitate the procedure,
we approximate the self-consistent scalar and vector potentials generated from the covariant 
energy density functional FSUGarnet\,\cite{Chen:2014mza}, by a  2-parameter Fermi 
(or Woods-Saxon) function, as illustrated in Fig.\ref{Figure1} for ${}^{208}$Pb. 

\begin{table}[ht]
\begin{tabular}{|c|c|c|c|c||c|c|c|}
 \hline\rule{0pt}{2.5ex}  
 Nucleus & $V_{\rm s}$ & $V_{\rm v}$ & $c$ & $a$ & 
 $\lambda_{\rm s}$ & $\lambda_{\rm v}$ & $\beta$ \\ 
 \hline\rule{0pt}{2.5ex}  
 $^{16}\mathrm{O}$    & 442.8 & 360.6 & 2.605 & 0.475 & 5.862 & 4.747 & 5.495 \\
 $^{40}\mathrm{Ca}$  & 464.4 & 378.8 & 3.565 & 0.621 & 8.417 & 6.821 & 5.748 \\
 $^{48}\mathrm{Ca}$  & 438.4 & 358.4 & 3.948 & 0.511 & 8.774 & 7.169 & 7.739 \\
 $^{68}\mathrm{Ni}$    & 420.8 & 345.4 & 4.513 & 0.520 & 9.639 & 7.886 & 8.686 \\
 $^{90}\mathrm{Zr}$    & 411.5 & 335.7 & 5.034 & 0.507 & 10.52 & 8.550 & 9.935 \\
 $^{132}\mathrm{Sn}$ & 400.1 & 329.2 & 5.803 & 0.498 & 11.77 & 9.670 & 11.67 \\
 $^{208}\mathrm{Pb}$ & 402.4 & 330.2 & 6.795 & 0.565 & 13.86 & 11.357 & 12.02 \\
 $^{302}\mathrm{Og}$ & 381.0 & 313.5 & 7.845 & 0.553 & 15.17 & 12.447 & 14.19 \\
 \hline
\end{tabular}
\caption{Optimal Woods-Saxon fit to the neutron scalar and vector potentials generated 
              by the covariant energy density functional FSUGarnet\,\cite{Chen:2014mza}. 
              The strength of the scalar and vector potential is given in MeV, and the
	      half-density radius and diffuseness in fm. The last three columns list the 
	      corresponding scaled, dimensionless parameters.}
 \label{Table1}
 \end{table} 
One then proceeds to scale both scalar and vector potentials as indicated in Eq.(\ref{Scaling}).
That is,
\begin{subequations}
\begin{eqnarray}
 & S(r) = \displaystyle{\frac{V_{\rm s}}{1+e^{(r-c)/a}}} \longrightarrow 
              \displaystyle{\sigma(x) = \frac{\lambda_{\rm s}}{1+e^{\,\beta(x-1)}}} \\
 & V(r) = \displaystyle{\frac{V_{\rm v}}{1+e^{(r-c)/a}}} \longrightarrow 
              \displaystyle{\omega(x) = \frac{\lambda_{\rm v}}{1+e^{\,\beta(x-1)}}},
\end{eqnarray}
\label{ScaledPots}
\end{subequations}
where $V_{\rm s}$ and $V_{\rm v}$ are the strengths of the scalar and vector potentials, respectively, 
$c$ is the half-density radius, and $a$ is the diffuseness parameter. Once scaled, the resulting potentials 
depend linearly on the dimensionless strengths $\lambda_{\rm s}$ and $\lambda_{\rm v}$, and non-linearly
on the dimensionless ratio $\beta\!\equiv\!c/a$. 

\begin{center}
\begin{figure}[ht]
\centering
 \includegraphics[width=0.5\textwidth]{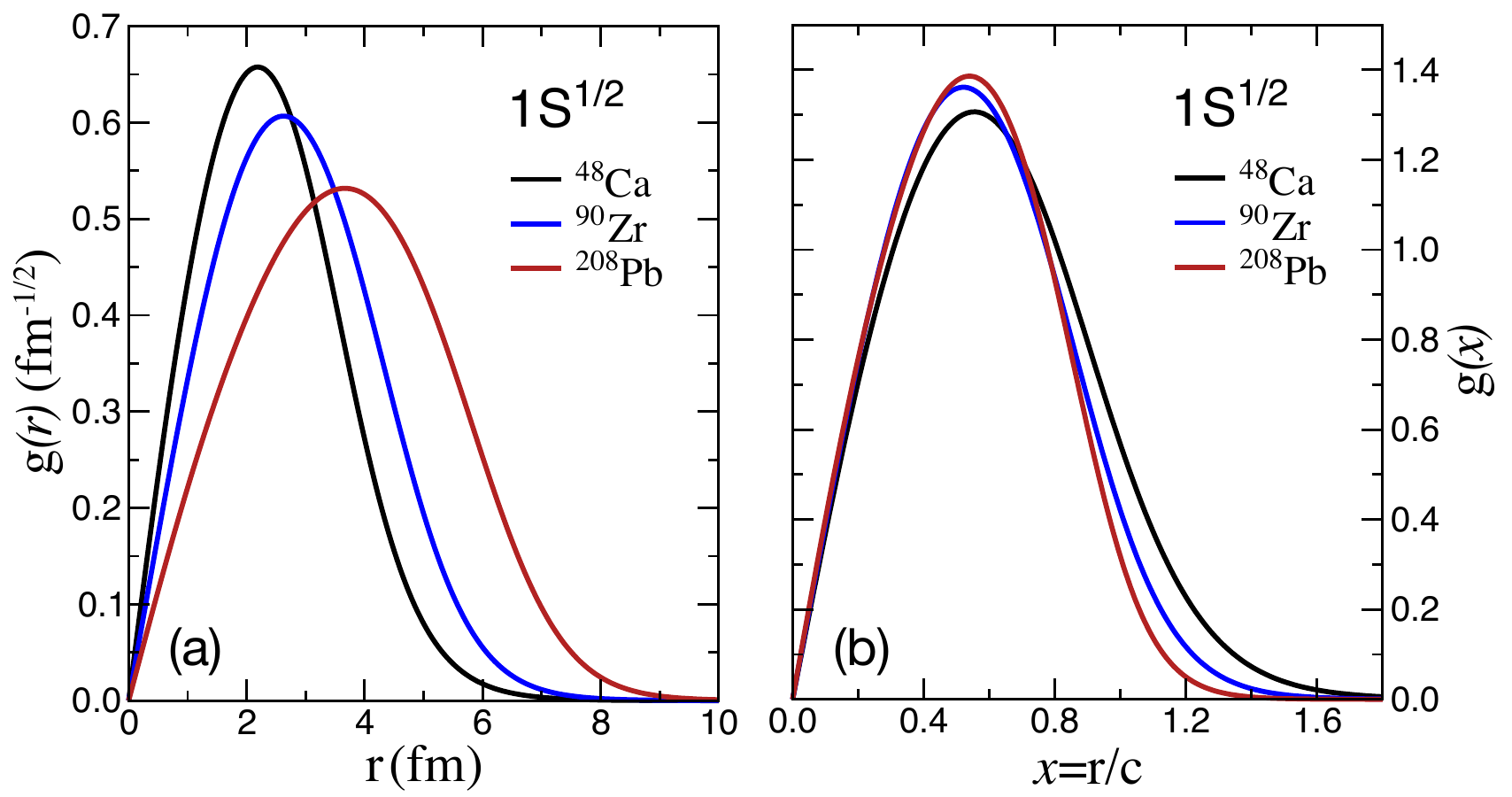}
 \caption{(a) The upper component of the ground-state Dirac orbital generated by solving the dimensionful 
                Dirac equation [Eq.(\ref{DiracEqn})] for ${}^{48}$Ca, ${}^{90}$Zr, and $^{208}$Pb. (b) Same as 
                (a) but now the Dirac orbitals are generated by solving the dimensionless Dirac equation 
                [Eq.(\ref{DiracEqns})] }
 \label{Figure2}
\end{figure}
\end{center}

Having defined the scaled potentials, one solves exactly the two-coupled differential equations given in 
Eqs.(\ref{DiracEqns}) by using a 4th-order Runge-Kutta algorithm for all nuclei listed in Table\,\ref{Table1},
with the exception of oganesson ($Z\!=\!118$ and $N\!=\!184$) that will be used later to test the efficacy 
of the emulator away from its training domain. As expected, the Dirac orbitals sharing the same quantum 
numbers are qualitatively similar. The simple rescaling implemented above makes the similarity between
Dirac orbitals quantitative, a fact that is instrumental in generating a universal reduced basis. The impact
of the scaling can be seen in Fig.\ref{Figure2} where the left-hand panel shows the ground state 
$1s^{1/2}(\kappa\!=\!-1)$ orbital generated by solving the original (dimensionful) Dirac equation. Instead,
the right-hand panel shows the same orbitals after rescaling the Dirac equation to a dimensionless form. 
As much of the $A$-dependence of the orbitals is mitigated by the scaling, the same reduced basis will be 
used to compute the single-energy spectrum of all spherical nuclei. Note that in the work reported in 
Ref.\,\cite{Giuliani:2022yna}, a reduced basis was created for each nucleus, and indeed, for each Dirac 
orbital. The availability of a universal reduced basis should further increase the efficiency of the emulator.

The last step in generating the (hopefully small) reduced basis invokes the Singular Value Decomposition 
(SVD) of an arbitrary matrix $A$. As in our earlier work\,\cite{Anderson:2022jhq}, we assemble all high
fidelity solutions with the same value of $\kappa$ into a rectangular $A$ matrix of dimension $m\times n$
and of rank $r$. That is, 
 \begin{equation}
   A=U \Sigma V^{\rm T} = u_{1}\sigma_{1}v_{1}^{T} + u_{2}\sigma_{2}v_{2}^{T} + \ldots + u_{r}\sigma_{r}v_{r}^{T} 
  \label{SVD}
 \end{equation}
where $U$ and $V$ are $m\times m$ and $n\times n$ orthogonal matrices, respectively, and $\Sigma$ is 
an $m\times n$ ``diagonal" matrix that encodes the rank of $A$, namely, the dimensionality of the vector 
space. As such, the matrix $A$ can be decomposed as a sum of rank-1 matrices, as in Eq.(\ref{SVD}), with 
the singular values $\sigma_{i}$ arranged in descending order of importance. This criteria identifies the most 
important elements of the orthonormal reduced basis. Moreover, the singular values estimate the error likely 
to incur from truncations of the reduced basis. Finally, note that each vector that we input into the matrix is of 
dimension $D\!=\!2X_{\rm max}$, where $X_{\rm max}$ is the number of spatial grid points required to 
describe the Dirac orbitals: $X_{\rm max}$ for the upper component and $X_{\rm max}$ for the lower
component.

\subsection{The Reduced Basis Method}
Having generated the reduced basis, we are now in a position to implement the reduced basis method
approach to calculate the eigenstates of the Dirac Hamiltonian. The main virtue of such an approach is
that whereas the solution of the high-fidelity model may be computationally demanding, the RBM transforms 
such challenging task into a diagonalization in a low-dimensional space spanned by the most important
elements of the reduced basis. Having taken advantage of the spherical symmetry of the problem, the
Dirac Hamiltonian matrix to be diagonalized---for each value of $\kappa$---is given by
 \begin{equation}
  \hat{H}(\kappa) =
 \begin{pmatrix}
   \mu-\sigma(x)+\omega(x)     & \displaystyle{-\frac{d}{dx}+\frac{\kappa}{x}}\\
   \displaystyle{\frac{d}{dx}+\frac{\kappa}{x}} & -\mu+\sigma(x)+\omega(x) 
  \end{pmatrix}.
  \label{DiracMatrix}
 \end{equation}
In turn, matrix elements of the Dirac Hamiltonian are given by the following expression
\begin{widetext}
 \begin{equation}
  \langle m|\hat{H}(\kappa)|n\rangle = \mathlarger{\int}_{0}^{\infty}
  \Big(g_{m\kappa}(x),f_{m\kappa}(x)\Big)
   \begin{pmatrix}
   \mu-\sigma(x)+\omega(x)     & \displaystyle{-\frac{d}{dx}+\frac{\kappa}{x}}\\
   \displaystyle{\frac{d}{dx}+\frac{\kappa}{x}} & -\mu+\sigma(x)+\omega(x) 
  \end{pmatrix}
     \begin{pmatrix}
      g_{n\kappa}(x) \vspace{3pt}\\ f_{n\kappa}(x)
     \end{pmatrix} dx,
 \label{DiracMatrixElem}
 \end{equation}
 \end{widetext}
where $(g_{n\kappa},f_{n\kappa})$ is the ${n\rm_{th}}$ element of the reduced basis in the
$\kappa$ sector.

\section{Results}
\label{Sec:Results}

In this section we present in detail the path travelled to reach our ultimate goal, namely, the 
construction of a universal reduced basis for the efficient calibration of covariant energy density 
functionals. We decided to illustrate some of our missteps to underscore some of the subtleties
behind the Dirac equation. Naturally, we started by assuming that the same approach employed 
in our previous publication to solve the Schr\"odinger equation\,\cite{Anderson:2022jhq} could 
also be used in the case of the Dirac equation. 

\subsection{The Naive Approach: Building a reduced basis \\ from only positive energy states}

As in the case of the Schr\"odinger equation, the high fidelity model is solved exactly for a given 
set of model parameters and then one collects all bound orbitals with a given quantum number. 
In the present case, we use the bound neutron orbitals generated by all nuclei listed in 
Table\,\ref{Table1}, with the exception of Oganesson. This collection of non-orthogonal states is 
then filtered through an SVD routine to generate an optimal orthonormal basis arranged in order of 
importance. The reduced basis is then obtained by keeping only those orbitals with a condition 
number of at least $10^{-3}$; that is, we keep orbitals whose singular value lies within 0.1\% of 
the largest one. Note that the $10^{-3}$ value provides a rule-of-thumb estimate that worked well 
in our previous study\,\cite{Anderson:2022jhq}.

\begin{figure}[ht]
 \centering
 \includegraphics[width=0.45\textwidth]{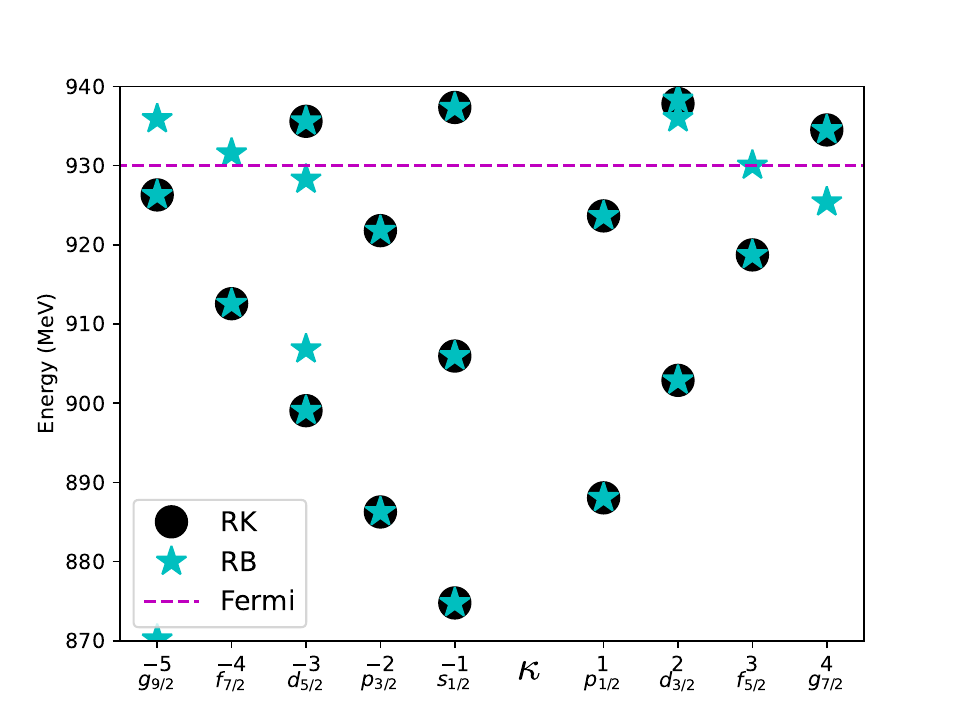}
 \caption{The bound single-particle spectrum of $^{90}$Zr. The black circles are the values obtained 
 from the numerical Runge-Kutta integration, while the blue stars are the values obtained from the RBM.
 The dashed line indicates the approximate location of the Fermi energy; all bound states below the 
 line are occupied where all the states above the line are empty.}
\label{Figure3}
\end{figure}

Following such a procedure and using ${}^{90}$Zr as an example, we show in Fig.\ref{Figure3} the entire 
single-neutron spectrum as generated by the high-fidelity model (black circles) as well as through
the reduced basis method (blue stars). Note that the energies $E$ reported here include the neutron rest mass 
of $M\!=\!940\,{\rm MeV}$, so the corresponding binding energies are simply given by $B\!=\!E\!-\!M\!<\!0$. 
Although the diagonalization of the matrix within the reduced basis correctly computes all bound state 
energies---both occupied and unoccupied---many unphysical states also appear. 

What may be the cause for the appearance of these unwanted ``ghost" energies? The answer to this
question lies in the nature of the Dirac spectrum. Indeed, the positive energy states of the Dirac 
equation---by themselves---do not form a complete basis. One of Dirac's great insights was the
discovery of negative-energy states that shortly after were identified as antiparticles. To shed light on 
how these ghost states emerge, we can regard the matrix diagonalization in the reduced basis as a 
variational approach that aims to select the $N$ optimal expansion parameters to describe the lowest
energy state for each quantum number. That is, we can write
 \begin{equation}
  |\psi\rangle = \sum_{n=1}^{N} a_{n}|n\rangle,
  \label{Expansion}
 \end{equation}
where $|\psi\rangle$ is an eigenstate of the Dirac Hamiltonian, $a_{n}$ are the expansion coefficients, and 
$|n\rangle$ is a member of the reduced basis introduced in Eq.(\ref{DiracMatrixElem}). Although the structure 
of the positive energy states is accurately imprinted in the reduced basis (e.g., the upper components are 
significantly larger than the corresponding lower components) the Dirac Hamiltonian ``knows" of the negative 
energy states. Hence, the extra ghost energies are an attempt to find the lowest energy solution. However, 
the attempt fails because bound states of negative energy can not be accurately expanded in terms of a 
reduced basis built from only positive energy states.

\subsection{The Comprehensive Approach: Building a reduced basis \\ from positive and negative energy states}

\begin{figure}[ht]
 \centering
 \includegraphics[width=0.47\textwidth]{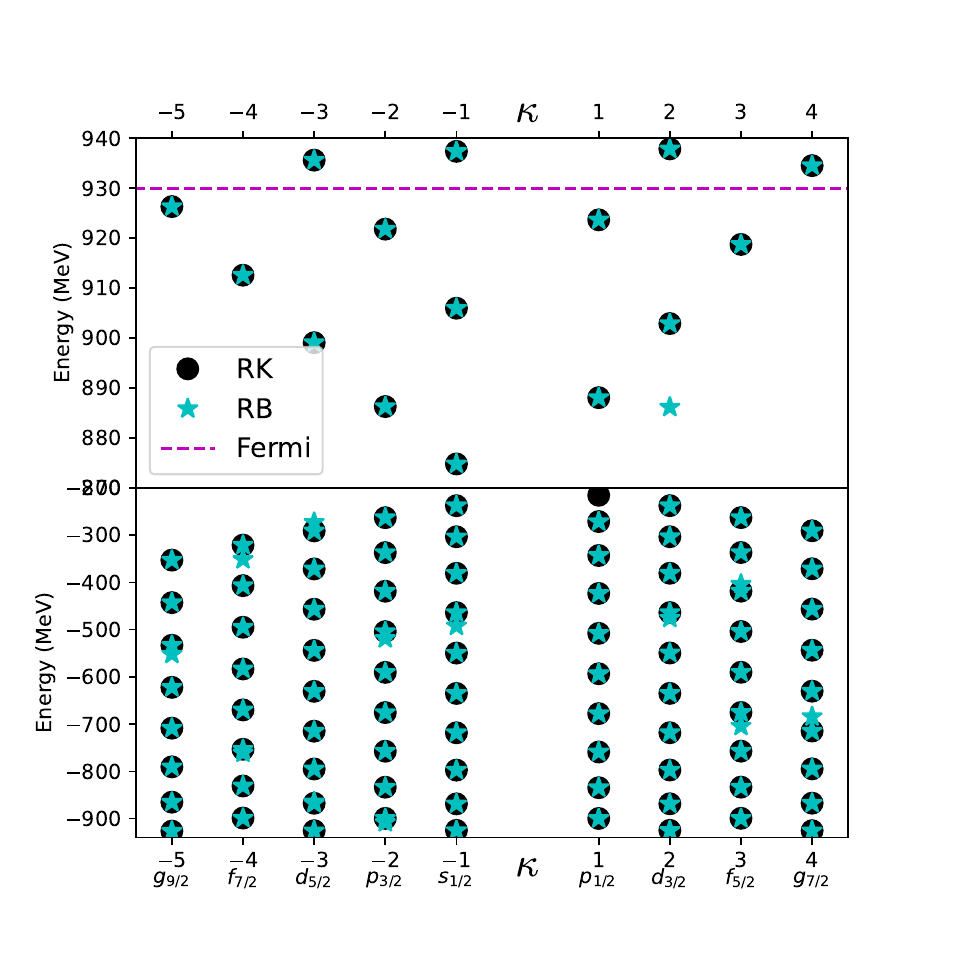}
 \caption{The entire---both positive and negative---bound single-particle spectrum of $^{90}$Zr.
 The black circles are the values obtained from the numerical Runge-Kutta integration, while the 
 blue stars are the values obtained from the RBM. The dashed line indicates the approximate location 
 of the Fermi energy; all positive-energy bound states below the line are occupied whereas all the 
 states above the line are empty.}
\label{Figure4}
\end{figure}

Faced with this dilemma, we now proceed to generate a new set of high-fidelity states, but this time by
including both positive and negative energy states. As before, we use bound neutron orbitals generated 
by all nuclei listed in Table\.,\ref{Table1}, with the exception of Oganesson. Although this involves a simple
change in sign of the energy $\epsilon$ appearing in Eqs.(\ref{DiracEqns}), the impact of this change is 
dramatic. Whereas in the case of the positive energy states the large scalar and vector potentials combine 
to produce a fairly weak central potential, the result is entirely different for the negative energy states. In 
particular, the central component of the effective Schr\"odinger equation, given by
 \begin{equation}
   U_{\rm c}(x)\approx-\sigma(x)+\frac{\epsilon}{\mu}\,\omega(x),
  \label{Schrodingerlike}
 \end{equation}
is fairly small for positive energy states because of the strong cancellation between large scalar and vector 
potentials. However, for negative energy states the strong cancellation is replaced by a strong enhancement, 
that leads to an explosion in the number of bound negative energy states. For example, in the case of ${}^{90}$Zr 
displayed in Fig.\ref{Figure4}, we find about 140 bound states---with only 15 of them having positive energy.
 
While the inclusion of the entire spectrum of positive and negative energy states eliminated most 
of the unwanted ghost energies, some problems still remain. First, eliminating the ghost energies
came at a very heavy price, as the number of basis states required to reach the necessary accuracy 
became enormous, reaching as high as 67 for some values of $\kappa$. Indeed, in order to eliminate 
all ghost energies, a very large number of basis states had to be retained, invalidating the idea 
of a ``reduced" basis. Second, even at this heavy cost, we observe some residual ghost energies. 
Although most of the ghost energies that appeared in Fig.\ref{Figure3} have been eliminated, a single 
$d^{3/2}$ ghost energy remains. Moreover, although most negative energy states are well reproduced, 
several unwanted ghost energies remain clearly visible. Note that no effort was made to explain the
reason for the few remaining ghosts.

\subsection{The Successful Approach: Building a reduced basis from \\ 
                    only positive energy states, but with a twist}

In an effort to overcome the problem of relying on an overly large reduced basis which mitigates but
does not entirely eliminate the appearance of ghost energies, we return to the original reduced basis 
built entirely from positive energy states. Although many unwanted ghost energies appear in Fig.\ref{Figure3}, 
the method seems to also reproduce the fifteen bound state energies obtained by solving the high-fidelity
model. Hence, the task at hand consists on filtering out the ghost energies from the true energies. To
do so, we implement the following two-fold prescription. First, we start by constructing a reduced basis 
from only positive energy state that we classify in terms of two quantum numbers $(n,\kappa)$, where 
$n$ is the ``principal" quantum number that accounts for the total number of nodes\,\cite{Giuliani:2022yna}. 
Second, once the Hamiltonian matrix is diagonalized within this reduced vector space, we only keep the 
eigenstate having the largest overlap with the most important element of the reduced basis. The most 
important element of the reduced basis, as quantified by its singular value, is guaranteed to have the 
same nodal structure as the exact eigenstate; the remaining elements of the reduced basis are simply 
in charge of performing the fine tuning.

\begin{figure}[ht]
 \centering
 \includegraphics[width=0.49\textwidth]{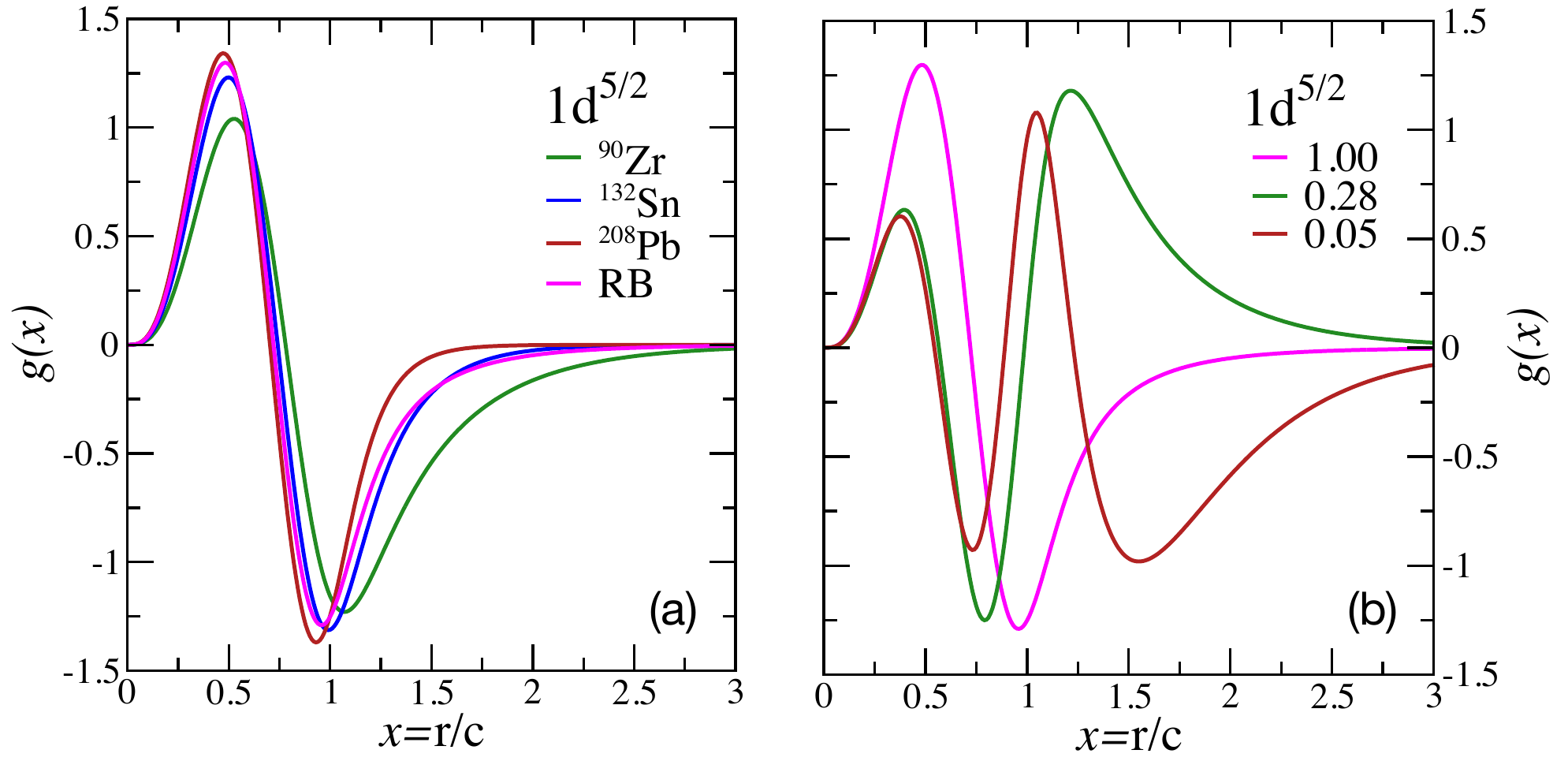}
 \caption{(a) Solutions of the high-fidelity model for the upper component of the $1d^{5/2}$ orbital
                    for ${}^{90}$Zr, ${}^{132}$Sn, and ${}^{208}$Pb. Also shown is the most important
                    element of the reduced basis. (b) The resulting orthonormal basis obtained by filtering 
                    the high-fidelity solution through an SVD algorithm.}
\label{Figure5}
\end{figure}

As an example on how we implement this procedure, we show in Fig.\,\ref{Figure5} how to generate the
$(n\!=\!1,\kappa\!=\!-3)$ Dirac orbital, namely, the $1d^{5/2}$ orbital containing one internal node. The 
left-hand panel in the figure displays high fidelity solutions for the upper (``large") component of the 
$1d^{5/2}$ orbital for ${}^{90}$Zr, ${}^{132}$Sn, and ${}^{208}$Pb. Note that this orbital is barely bound 
in ${}^{90}$Zr; see Fig.\ref{Figure4}. In turn, the right-hand panel in Fig.\ref{Figure5} displays the optimal 
orthonormal reduced basis generated by SVD. As anticipated, the basis with the largest condition number 
has the same structure as the high-fidelity states. To underscore the similarity, we added this most important 
reduced-basis basis state to the left-hand panel. 
\begin{figure}[ht]
 \vspace{5pt}
 \centering
 \includegraphics[width=0.35\textwidth]{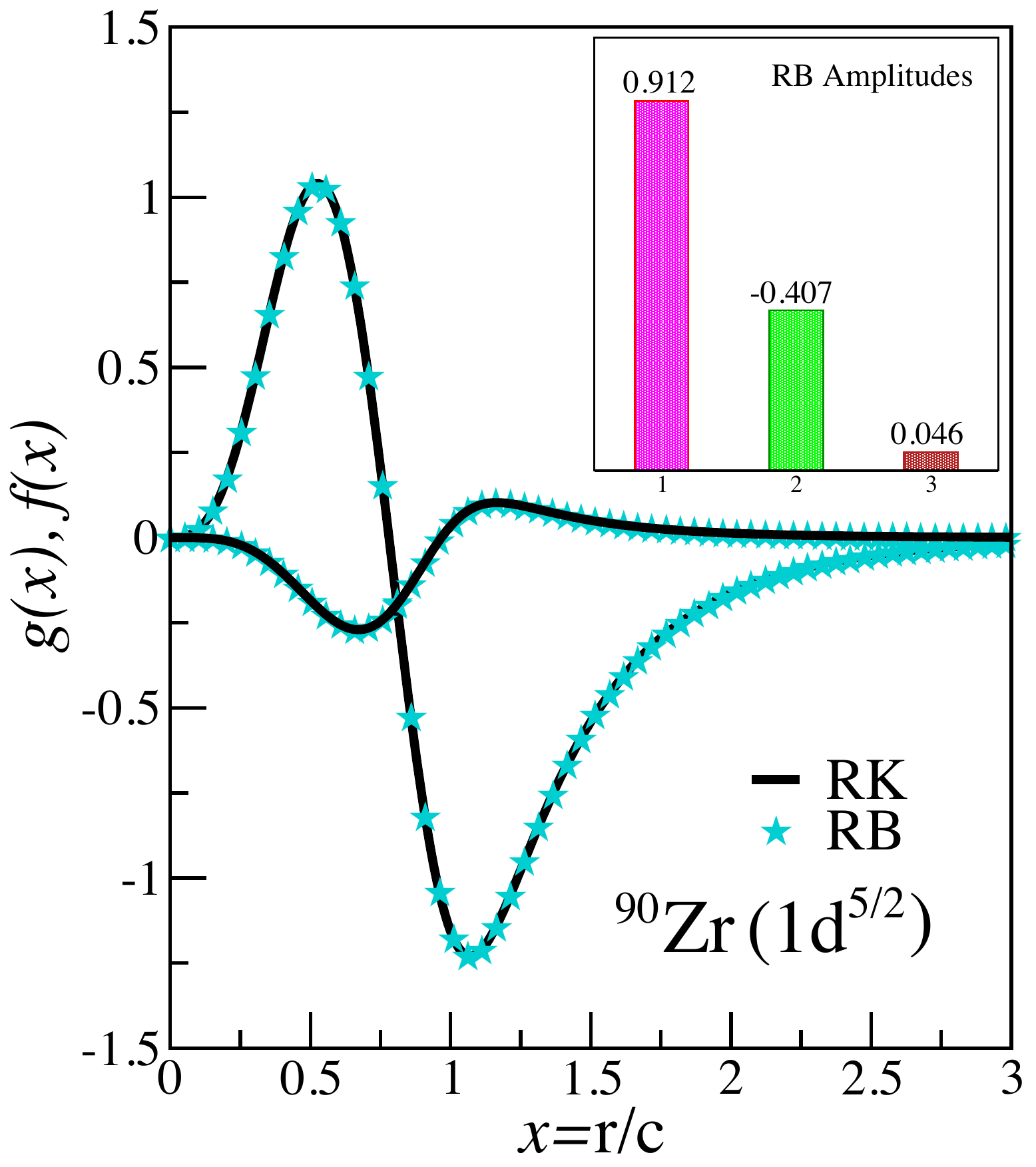}
 \caption{Upper (``large") and lower (``small") components of the $1d^{5/2}$ Dirac orbital in ${}^{90}$Zr, 
               as obtained from the numerical solution (RK) and from using a reduced basis (RB) method. The 
               inset shows the projection of the Dirac orbital into the three reduced basis states.}
\label{Figure6}
\end{figure}

\begin{widetext}
\begin{figure*}[ht]
 \centering
 \includegraphics[width=0.99\textwidth]{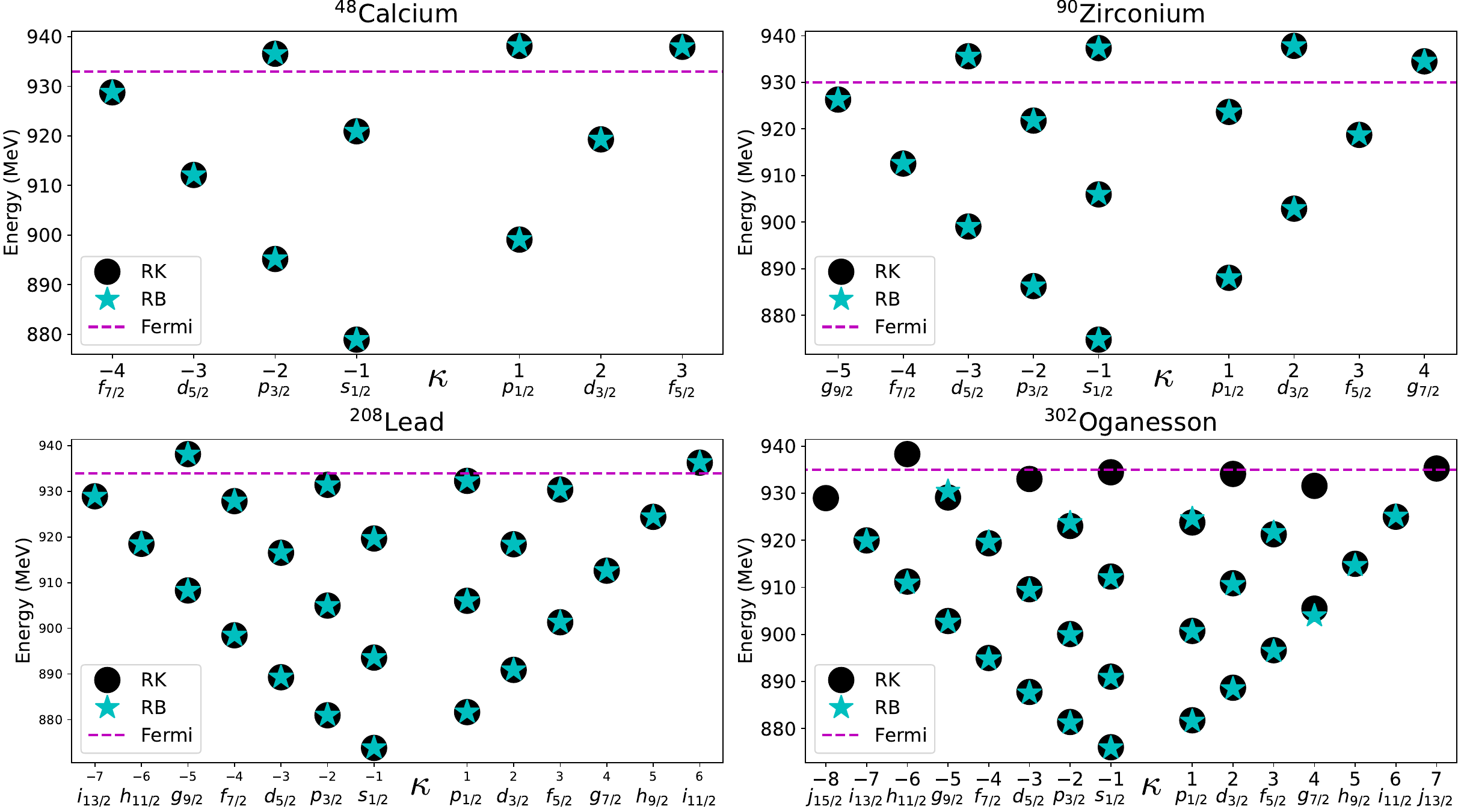}
 \caption{The bound, positive-energy single-particle spectrum of $^{48}$Ca $^{90}$Zr, $^{208}$Pb,
 and $^{302}$Og. The black circles denote the values obtained from the numerical RK integration, 
 while the blue stars are the values obtained from the RBM. The dashed line indicates the approximate 
 location of the Fermi energy; all bound states below the line are occupied where all the states above 
 the line are empty. The uncomputed energies in $^{302}$Og are a result 
 of not having input functions to create a basis for those states.}
\label{Figure7}
\end{figure*}
\end{widetext}

Once the reduced basis has been obtained, one proceeds to diagonalize the ``tiny" $3\!\times\!3$ Dirac 
Hamiltonian matrix. The result of the diagonalization procedure is displayed in Fig.\ref{Figure6}, which 
shows results for both the large upper $g(x)$ and small lower $f(x)$ components. The high fidelity solution 
obtained from using the Runge-Kutta algorithm is shown by the black solid line while the one obtained 
from the matrix diagonalization is depicted with the blue stars. Shown in the inset are the reduced-basis 
amplitudes as defined in Eq.(\ref{Expansion}). As expected, the first element of the reduced basis carries 
the largest weight, the second element provides most of the fine tuning, and the third element can be largely 
ignored. This procedure is applied to all nuclei in Table\,\ref{Table1} and the results for four of them are 
displayed in Fig.\ref{Figure7}. Note that ${}^{302}\,{\rm Og}$ was not included in the generation of the 
reduced basis, so its inclusion tests the performance of the model in extrapolating away from its training 
domain. Also note that some of the bound states in ${}^{302}\,{\rm Og}$ do not contain an RB counterpart
because of the obvious limitations of a reduced basis built only from lighter nuclei. Finally, the optimal 
reduced basis to be used in future calculations will benefit from the addition of ${}^{302}\,{\rm Og}$.

We now summarize the advantages of implementing the two-fold procedure described above. By
concentrating on orbitals classified according to the two quantum numbers $(n,\kappa)$, one can 
generate a small and efficient reduced basis for each channel. By virtue of the small dimensionality 
of the reduced basis, the diagonalization of the Dirac matrix is extremely fast. From the few eigenstates 
that emerge from the diagonalization, we are only interested in selecting one, namely, the one with 
the largest overlap with the most important element of the reduced basis. For all the bound orbitals
below the Fermi surface that we have examined, the largest overlap exceeds 90\%. Moreover, after
rescaling the Dirac Hamiltonian according to Eq.(\ref{DiracEqns}), the reduced basis is ``universal",
in the sense that it remains unchanged as one goes from ${}^{16}\,{\rm O}$ to ${}^{302}\,{\rm Og}$.

We conclude this section by making contact with our earlier publication\,\cite{Anderson:2022jhq}.
While the framework just described, namely, an emulator created with a distinct reduced basis for 
each set of $(n,\kappa)$ was successful, we realized that the filtering algorithm used in our earlier 
work consisting of a single basis for each $\kappa$ can also work, as long as we ensure that the 
selected eigenvectors continue to have the largest overlap with the most relevant element of the 
reduced basis. Although the resulting Dirac matrix is larger, and thus takes longer to diagonalize, 
the advantage of creating only a single matrix for each $\kappa$ results in an additional speedup 
factor of about 2.5, relative to the version using multiple smaller matrices for each $\kappa$. 

\subsection{Linearizing the Potential}

The emulator that we have constructed so far speeds up the calculation of Dirac orbitals by nearly a 
factor of 150. While not insignificant, even such a considerable speedup factor often strains our 
computational resources. In particular, Bayesian calibration of covariant energy density 
functionals demand performing this sort of calculations thousands (if not millions) of times. As already
mentioned, the diagonalization of the Dirac matrix in small vector spaces is very fast, so most of the 
computational time is spent computing the matrix elements given in Eq.(\ref{DiracMatrixElem}). So if
this bottleneck could be eliminated by precomputing all matrix elements, one would gain an additional
advantage. 

However, given that the scaled Woods-Saxon potentials are non-linear in the parameter $\beta\!=\!c/\!a$ 
listed in Table\,\ref{Table1}, one would have to precompute matrix elements for each individual nucleus, 
thereby breaking the ``universality" of the entire approach. Yet, one could mitigate this by linearizing the 
potential, thereby improving the dimensionality reduction of the problem. For example, one could write the 
dimensionless Woods-Saxon form given in Eq.(\ref{ScaledPots}) as
follows:
\begin{equation}
 \mathlarger{u}(x;\beta) = \frac{1}{1+e^{\,\beta(x-1)}} \!\approx\!\sum_{q=1}^{Q} b_{q}(\beta)\mathlarger{u}_{q}(x),
 \label{EIM}
\end{equation}
where the functions $\mathlarger{u}_{q}(x)$ may be obtained following a procedure similar to the one adopted to 
construct the reduced basis. That is, one inputs into SVD a set of dimensionless Woods-Saxon forms spanning 
the range of values of $\beta$ given in Table\,\ref{Table1}. Because SVD generates an optimal set of orthonormal 
functions, the nucleus-specific coefficients may be readily computed. That is,
\begin{equation}
 b_{q}(\beta) = \int_{0}^{\infty} \mathlarger{u}(x;\beta)\mathlarger{u}_{q}(x) dx.
 \label{WSExpansion}
\end{equation}
Hence, by linearizing the potential, the information specific to a given nucleus is encoded in the coefficients $b_{q}(\beta)$, 
but matrix elements of $\mathlarger{u}_{q}(x)$ between reduced basis states become independent of the nucleus under 
consideration. For example, matrix elements of the dimensionless vector potential 
$\omega(x;\beta)$ may now be written according to Eq.(\ref{DiracMatrixElem}) as follows:
 \begin{equation}
 \begin{aligned}
  \langle m\kappa|\hat{\omega}(\beta)|n\kappa\rangle = 
  & \lambda_{\rm v}(\beta)\langle m\kappa|\hat{\mathlarger{u}}(\beta) |n\kappa\rangle \\ = 
  & \lambda_{\rm v}(\beta)\sum_{q=1}^{Q} b_{q}(\beta)\langle m\kappa|\hat{\mathlarger{u}}_{q}|n\kappa\rangle.  
  \end{aligned}
  \label{VectorMatrixElem}
 \end{equation}
As such, the entire nuclear dependence is contained in the strength of the potential $\lambda_{\rm v}(\beta)$ 
and in the $b_{q}(\beta)$ coefficients. The remaining matrix elements, given by
\begin{equation}  
  \langle m\kappa|\hat{\mathlarger{u}}_{q}|n\kappa\rangle\!=\!\!
  \int_{0}^{\infty}\!\!\Big(g_{m\kappa}(x)g_{n\kappa}(x)\!+\!f_{m\kappa}(x)f_{n\kappa}(x)\Big)\mathlarger{u}_{q}(x)dx,
  \label{VectorMatrixElem}
 \end{equation}
are ``universal", so they can be precomputed offline and then stored for later use. 

Once the potentials have been linearized and the relevant matrix elements computed and stored, all that remains is to build 
the matrix elements and diagonalize the various Dirac matrices in the reduced vector spaces. Using three expansion coefficients 
to linearize the potentials, the resulting emulator is about 
12,000 times faster than the numerical RK integration method, and almost 100 times faster than using the reduced basis method 
in its original form. Although the linearization introduces a small additional error, the error is not visible when plotted against the 
exact numerical result. Hence the gains in computational expediency afforded by the linearization more than compensate for the 
slightly larger error.  

\section{Conclusions and Outlook}
\label{Sec:Conclusions}

Emulators of high-fidelity models provide an efficient and accurate way to solve problems that often require long, computationally 
expensive numerical calculations. The Bayesian calibration of covariant energy density functionals falls into this category as it
involves computing several nuclear properties for a variety of nuclei thousands of times. As part of the calibration protocol, one 
must solve the Dirac equation for all the occupied orbitals for all nuclei under consideration. In the high-fidelity model this is done
by using the numerical RK integration method. Moreover, because the potentials generating the bound states depend themselves
on the occupied orbitals, the problem becomes non-linear and must then be solved self-consistently, which exacerbates the 
computational demands. 

In this paper we have extended our earlier work on the non-relativistic Schr\"odinger problem to the relativistic domain. As in the 
non-relativistic case, we built an efficient and accurate emulator by adopting a universal reduced basis capable of generating the 
entire single-particle spectrum of a variety of spherical nuclei without the need to redefine the basis. However, in contrast to the 
Schr\"odinger case, generating the relativistic reduced basis was complicated by the unavoidable existence of negative energy 
states. Although the problem was ultimately solved, we decided to document some of our tribulations, both because of the subtleties 
of the Dirac equation and to alert the reader of some our missteps. Ultimately, generating an efficient and robust basis required a 
few simple modifications relative to the strategy adopted in our earlier paper\,\cite{Anderson:2022jhq}. 

The successful reduced basis was built by generating high-fidelity solutions of the Dirac equations over a relative wide range of
angular momentum $\kappa$. These solutions were then classified according to the quantum number $\kappa$. 
For each value of $\kappa$, an orthonormal reduced basis was 
generated using a singular value decomposition. Next, matrix elements of the Dirac Hamiltonian were generated for each 
$\kappa$ and then the matrix was brought to diagonal form. From all the eigenvectors, one selects those having
the largest overlap with one of the three most important reduced basis states, namely, with the ones having the largest singular values. 
This guarantees that all unwanted ``ghosts" states that contaminated the spectrum disappear. We note that for all Dirac orbitals
and for all nuclei, accurate results were obtained by diagonalizing at most a $12\!\times\!12$ matrix. Moreover, in most of the 
cases, the most important element of the reduced basis carried at least 90\% of the weight. With this implementation, we saw
speed ups of about a factor of 150, relative to the high-fidelity model. 

Whereas a factor of 150 is significant, we identified the evaluation of the matrix elements as the most serious bottleneck in the 
problem. This bottleneck develops because the scalar and vector potentials depend non-linearly on the parameter $\beta$, 
which is different for each nucleus. We mitigated the problem by invoking a linearization scheme. The linearization 
improves greatly the dimensionality reduction of the problem by expanding the non-linear potential in a set of basis states
that are independent of the nucleus and a corresponding set of coefficients that carry the nuclear information. In this manner,
matrix elements of the nucleus-independent basis states may be precomputed and stored, leading to an even greater 
computational expediency. Indeed, precomputing matrix elements of the potential in the small reduced vector spaces 
yields an additional speed-up factor of 100 relative to the RBM in its original form. That is, by implementing these methods the 
calculations can be performed thousands of times faster, without sacrificing accuracy.

In the future, we aim to mitigate the last remaining bottleneck associated with the iterative procedure required to solve
the mean-field problem self-consistently. Indeed, besides a well-motivated reduced basis that captures the 
essential physics, one also requires an efficient framework to solve the underlying set of dynamical equations. For a
system of linear differential equation as the one considered here, the solution simply involves direct diagonalization of 
the Hamiltonian matrix in the reduced-basis space. However, for the self-consistent solution of the mean-field problem, 
one needs to solve a non-linear set of differential equations. In this case, rather than solving the self-consistent problem
by iteration, one invokes the Galerkin projection approach, an enormously efficient framework that involves solving once
a non-linear set of algebraic equations\,\cite{Quarteroni:2015}. Although without the benefit of a universal reduced basis,
significant progress along this direction has already been accomplished in Ref.\,\cite{Giuliani:2022yna}. 

\begin{acknowledgments}\vspace{-10pt}
 We are grateful to Pablo Giuliani for many useful and insightful discussions. This material is based upon work supported 
 by the U.S. Department of Energy Office of Science, Office of Nuclear Physics under Award Number DE-FG02-92ER40750 and 
 under the STREAMLINE collaboration award DE-SC0024646.
\end{acknowledgments} 

\bibliography{./ReferencesJP}

\begin{thebibliography}{47}
\expandafter\ifx\csname natexlab\endcsname\relax\def\natexlab#1{#1}\fi
\expandafter\ifx\csname bibnamefont\endcsname\relax
  \def\bibnamefont#1{#1}\fi
\expandafter\ifx\csname bibfnamefont\endcsname\relax
  \def\bibfnamefont#1{#1}\fi
\expandafter\ifx\csname citenamefont\endcsname\relax
  \def\citenamefont#1{#1}\fi
\expandafter\ifx\csname url\endcsname\relax
  \def\url#1{\texttt{#1}}\fi
\expandafter\ifx\csname urlprefix\endcsname\relax\def\urlprefix{URL }\fi
\providecommand{\bibinfo}[2]{#2}
\providecommand{\eprint}[2][]{\url{#2}}

\bibitem[{\citenamefont{PRA-Editors}(2011)}]{PhysRevA.83.040001}
\bibinfo{author}{\bibnamefont{PRA-Editors}}, \bibinfo{journal}{Phys. Rev. A}
  \textbf{\bibinfo{volume}{83}}, \bibinfo{pages}{040001}
  (\bibinfo{year}{2011}).

\bibitem[{\citenamefont{Reinhard and Nazarewicz}(2010)}]{Reinhard:2010wz}
\bibinfo{author}{\bibfnamefont{P.-G.} \bibnamefont{Reinhard}} \bibnamefont{and}
  \bibinfo{author}{\bibfnamefont{W.}~\bibnamefont{Nazarewicz}},
  \bibinfo{journal}{Phys. Rev.} \textbf{\bibinfo{volume}{C81}},
  \bibinfo{pages}{051303(R)} (\bibinfo{year}{2010}).

\bibitem[{\citenamefont{Kortelainen et~al.}(2010)\citenamefont{Kortelainen,
  Lesinski, More, Nazarewicz, Sarich et~al.}}]{Kortelainen:2010hv}
\bibinfo{author}{\bibfnamefont{M.}~\bibnamefont{Kortelainen}},
  \bibinfo{author}{\bibfnamefont{T.}~\bibnamefont{Lesinski}},
  \bibinfo{author}{\bibfnamefont{J.}~\bibnamefont{More}},
  \bibinfo{author}{\bibfnamefont{W.}~\bibnamefont{Nazarewicz}},
  \bibinfo{author}{\bibfnamefont{J.}~\bibnamefont{Sarich}},
  \bibnamefont{et~al.}, \bibinfo{journal}{Phys. Rev.}
  \textbf{\bibinfo{volume}{C82}}, \bibinfo{pages}{024313}
  (\bibinfo{year}{2010}).

\bibitem[{\citenamefont{Fattoyev and Piekarewicz}(2011)}]{Fattoyev:2011ns}
\bibinfo{author}{\bibfnamefont{F.}~\bibnamefont{Fattoyev}} \bibnamefont{and}
  \bibinfo{author}{\bibfnamefont{J.}~\bibnamefont{Piekarewicz}},
  \bibinfo{journal}{Phys. Rev.} \textbf{\bibinfo{volume}{C84}},
  \bibinfo{pages}{064302} (\bibinfo{year}{2011}).

\bibitem[{\citenamefont{Reinhard et~al.}(2013)\citenamefont{Reinhard,
  Piekarewicz, Nazarewicz, Agrawal, Paar et~al.}}]{Reinhard:2013fpa}
\bibinfo{author}{\bibfnamefont{P.-G.} \bibnamefont{Reinhard}},
  \bibinfo{author}{\bibfnamefont{J.}~\bibnamefont{Piekarewicz}},
  \bibinfo{author}{\bibfnamefont{W.}~\bibnamefont{Nazarewicz}},
  \bibinfo{author}{\bibfnamefont{B.}~\bibnamefont{Agrawal}},
  \bibinfo{author}{\bibfnamefont{N.}~\bibnamefont{Paar}}, \bibnamefont{et~al.},
  \bibinfo{journal}{Phys.Rev.} \textbf{\bibinfo{volume}{C88}},
  \bibinfo{pages}{034325} (\bibinfo{year}{2013}).

\bibitem[{\citenamefont{Dobaczewski et~al.}(2014)\citenamefont{Dobaczewski,
  Nazarewicz, and Reinhard}}]{Dobaczewski:2014jga}
\bibinfo{author}{\bibfnamefont{J.}~\bibnamefont{Dobaczewski}},
  \bibinfo{author}{\bibfnamefont{W.}~\bibnamefont{Nazarewicz}},
  \bibnamefont{and} \bibinfo{author}{\bibfnamefont{P.-G.}
  \bibnamefont{Reinhard}}, \bibinfo{journal}{J. Phys.}
  \textbf{\bibinfo{volume}{G41}}, \bibinfo{pages}{074001}
  (\bibinfo{year}{2014}).

\bibitem[{\citenamefont{Piekarewicz et~al.}(2015)\citenamefont{Piekarewicz,
  Chen, and Fattoyev}}]{Piekarewicz:2014kza}
\bibinfo{author}{\bibfnamefont{J.}~\bibnamefont{Piekarewicz}},
  \bibinfo{author}{\bibfnamefont{W.-C.} \bibnamefont{Chen}}, \bibnamefont{and}
  \bibinfo{author}{\bibfnamefont{F.}~\bibnamefont{Fattoyev}},
  \bibinfo{journal}{J. Phys.} \textbf{\bibinfo{volume}{G42}},
  \bibinfo{pages}{034018} (\bibinfo{year}{2015}).

\bibitem[{\citenamefont{Chen and Piekarewicz}(2014)}]{Chen:2014sca}
\bibinfo{author}{\bibfnamefont{W.-C.} \bibnamefont{Chen}} \bibnamefont{and}
  \bibinfo{author}{\bibfnamefont{J.}~\bibnamefont{Piekarewicz}},
  \bibinfo{journal}{Phys. Rev.} \textbf{\bibinfo{volume}{C90}},
  \bibinfo{pages}{044305} (\bibinfo{year}{2014}).

\bibitem[{\citenamefont{Wesolowski et~al.}(2016)\citenamefont{Wesolowski, Klco,
  Furnstahl, Phillips, and Thapaliya}}]{Wesolowski:2015fqa}
\bibinfo{author}{\bibfnamefont{S.}~\bibnamefont{Wesolowski}},
  \bibinfo{author}{\bibfnamefont{N.}~\bibnamefont{Klco}},
  \bibinfo{author}{\bibfnamefont{R.}~\bibnamefont{Furnstahl}},
  \bibinfo{author}{\bibfnamefont{D.}~\bibnamefont{Phillips}}, \bibnamefont{and}
  \bibinfo{author}{\bibfnamefont{A.}~\bibnamefont{Thapaliya}},
  \bibinfo{journal}{J. Phys. G} \textbf{\bibinfo{volume}{43}},
  \bibinfo{pages}{074001} (\bibinfo{year}{2016}).

\bibitem[{\citenamefont{Yuksel et~al.}(2019)\citenamefont{Yuksel, Marketin, and
  Paar}}]{Yuksel:2019dnp}
\bibinfo{author}{\bibfnamefont{E.}~\bibnamefont{Yuksel}},
  \bibinfo{author}{\bibfnamefont{T.}~\bibnamefont{Marketin}}, \bibnamefont{and}
  \bibinfo{author}{\bibfnamefont{N.}~\bibnamefont{Paar}},
  \bibinfo{journal}{Phys. Rev.} \textbf{\bibinfo{volume}{C99}},
  \bibinfo{pages}{034318} (\bibinfo{year}{2019}).

\bibitem[{\citenamefont{Drischler
  et~al.}(2020{\natexlab{a}})\citenamefont{Drischler, Furnstahl, Melendez, and
  Phillips}}]{Drischler:2020hwi}
\bibinfo{author}{\bibfnamefont{C.}~\bibnamefont{Drischler}},
  \bibinfo{author}{\bibfnamefont{R.}~\bibnamefont{Furnstahl}},
  \bibinfo{author}{\bibfnamefont{J.}~\bibnamefont{Melendez}}, \bibnamefont{and}
  \bibinfo{author}{\bibfnamefont{D.}~\bibnamefont{Phillips}},
  \bibinfo{journal}{Phys. Rev. Lett.} \textbf{\bibinfo{volume}{125}},
  \bibinfo{pages}{202702} (\bibinfo{year}{2020}{\natexlab{a}}).

\bibitem[{\citenamefont{Drischler
  et~al.}(2020{\natexlab{b}})\citenamefont{Drischler, Melendez, Furnstahl, and
  Phillips}}]{Drischler:2020yad}
\bibinfo{author}{\bibfnamefont{C.}~\bibnamefont{Drischler}},
  \bibinfo{author}{\bibfnamefont{J.~A.} \bibnamefont{Melendez}},
  \bibinfo{author}{\bibfnamefont{R.~J.} \bibnamefont{Furnstahl}},
  \bibnamefont{and} \bibinfo{author}{\bibfnamefont{D.~R.}
  \bibnamefont{Phillips}}, \bibinfo{journal}{Phys. Rev. C}
  \textbf{\bibinfo{volume}{102}}, \bibinfo{pages}{054315}
  (\bibinfo{year}{2020}{\natexlab{b}}).

\bibitem[{\citenamefont{Furnstahl et~al.}(2020)\citenamefont{Furnstahl, Garcia,
  Millican, and Zhang}}]{Furnstahl:2020abp}
\bibinfo{author}{\bibfnamefont{R.~J.} \bibnamefont{Furnstahl}},
  \bibinfo{author}{\bibfnamefont{A.~J.} \bibnamefont{Garcia}},
  \bibinfo{author}{\bibfnamefont{P.~J.} \bibnamefont{Millican}},
  \bibnamefont{and} \bibinfo{author}{\bibfnamefont{X.}~\bibnamefont{Zhang}},
  \bibinfo{journal}{Phys. Lett. B} \textbf{\bibinfo{volume}{809}},
  \bibinfo{pages}{135719} (\bibinfo{year}{2020}).

\bibitem[{\citenamefont{Athanassopoulos
  et~al.}(2004)\citenamefont{Athanassopoulos, Mavrommatis, Gernoth, and
  Clark}}]{Athanassopoulos:2003qe}
\bibinfo{author}{\bibfnamefont{S.}~\bibnamefont{Athanassopoulos}},
  \bibinfo{author}{\bibfnamefont{E.}~\bibnamefont{Mavrommatis}},
  \bibinfo{author}{\bibfnamefont{K.~A.} \bibnamefont{Gernoth}},
  \bibnamefont{and} \bibinfo{author}{\bibfnamefont{J.~W.} \bibnamefont{Clark}},
  \bibinfo{journal}{Nucl. Phys.} \textbf{\bibinfo{volume}{A743}},
  \bibinfo{pages}{222} (\bibinfo{year}{2004}).

\bibitem[{\citenamefont{Akkoyun et~al.}(2013)\citenamefont{Akkoyun, Bayram,
  Kara, and Sinan}}]{Akkoyun:2012yf}
\bibinfo{author}{\bibfnamefont{S.}~\bibnamefont{Akkoyun}},
  \bibinfo{author}{\bibfnamefont{T.}~\bibnamefont{Bayram}},
  \bibinfo{author}{\bibfnamefont{S.~O.} \bibnamefont{Kara}}, \bibnamefont{and}
  \bibinfo{author}{\bibfnamefont{A.}~\bibnamefont{Sinan}}, \bibinfo{journal}{J.
  Phys.} \textbf{\bibinfo{volume}{G40}}, \bibinfo{pages}{055106}
  (\bibinfo{year}{2013}).

\bibitem[{\citenamefont{Bayram et~al.}(2014)\citenamefont{Bayram, Akkoyun, and
  Kara}}]{Bayram:2013hi}
\bibinfo{author}{\bibfnamefont{T.}~\bibnamefont{Bayram}},
  \bibinfo{author}{\bibfnamefont{S.}~\bibnamefont{Akkoyun}}, \bibnamefont{and}
  \bibinfo{author}{\bibfnamefont{S.~O.} \bibnamefont{Kara}},
  \bibinfo{journal}{Annals of Nuclear Energy} \textbf{\bibinfo{volume}{63}},
  \bibinfo{pages}{172} (\bibinfo{year}{2014}).

\bibitem[{\citenamefont{Gernoth et~al.}(1993)\citenamefont{Gernoth, Clark,
  Prater, and Bohr}}]{Gernoth:1993}
\bibinfo{author}{\bibfnamefont{K.}~\bibnamefont{Gernoth}},
  \bibinfo{author}{\bibfnamefont{J.}~\bibnamefont{Clark}},
  \bibinfo{author}{\bibfnamefont{J.}~\bibnamefont{Prater}}, \bibnamefont{and}
  \bibinfo{author}{\bibfnamefont{H.}~\bibnamefont{Bohr}},
  \bibinfo{journal}{Phys. Lett. B} \textbf{\bibinfo{volume}{300}},
  \bibinfo{pages}{1} (\bibinfo{year}{1993}).

\bibitem[{\citenamefont{Utama et~al.}(2016{\natexlab{a}})\citenamefont{Utama,
  Piekarewicz, and Prosper}}]{Utama:2015hva}
\bibinfo{author}{\bibfnamefont{R.}~\bibnamefont{Utama}},
  \bibinfo{author}{\bibfnamefont{J.}~\bibnamefont{Piekarewicz}},
  \bibnamefont{and} \bibinfo{author}{\bibfnamefont{H.~B.}
  \bibnamefont{Prosper}}, \bibinfo{journal}{Phys. Rev.}
  \textbf{\bibinfo{volume}{C93}}, \bibinfo{pages}{014311}
  (\bibinfo{year}{2016}{\natexlab{a}}).

\bibitem[{\citenamefont{Utama et~al.}(2016{\natexlab{b}})\citenamefont{Utama,
  Chen, and Piekarewicz}}]{Utama:2016rad}
\bibinfo{author}{\bibfnamefont{R.}~\bibnamefont{Utama}},
  \bibinfo{author}{\bibfnamefont{W.-C.} \bibnamefont{Chen}}, \bibnamefont{and}
  \bibinfo{author}{\bibfnamefont{J.}~\bibnamefont{Piekarewicz}},
  \bibinfo{journal}{J. Phys.} \textbf{\bibinfo{volume}{G}},
  \bibinfo{pages}{014311} (\bibinfo{year}{2016}{\natexlab{b}}).

\bibitem[{\citenamefont{Utama and Piekarewicz}(2018)}]{Utama:2017ytc}
\bibinfo{author}{\bibfnamefont{R.}~\bibnamefont{Utama}} \bibnamefont{and}
  \bibinfo{author}{\bibfnamefont{J.}~\bibnamefont{Piekarewicz}},
  \bibinfo{journal}{Phys. Rev.} \textbf{\bibinfo{volume}{C97}},
  \bibinfo{pages}{014306} (\bibinfo{year}{2018}).

\bibitem[{\citenamefont{Neufcourt et~al.}(2019)\citenamefont{Neufcourt, Cao,
  Nazarewicz, Olsen, and Viens}}]{Neufcourt:2019qvd}
\bibinfo{author}{\bibfnamefont{L.}~\bibnamefont{Neufcourt}},
  \bibinfo{author}{\bibfnamefont{Y.}~\bibnamefont{Cao}},
  \bibinfo{author}{\bibfnamefont{W.}~\bibnamefont{Nazarewicz}},
  \bibinfo{author}{\bibfnamefont{E.}~\bibnamefont{Olsen}}, \bibnamefont{and}
  \bibinfo{author}{\bibfnamefont{F.}~\bibnamefont{Viens}},
  \bibinfo{journal}{Phys. Rev. Lett.} \textbf{\bibinfo{volume}{122}},
  \bibinfo{pages}{062502} (\bibinfo{year}{2019}).

\bibitem[{\citenamefont{Neufcourt et~al.}(2020)\citenamefont{Neufcourt, Cao,
  Giuliani, Nazarewicz, Olsen, and Tarasov}}]{Neufcourt:2020nme}
\bibinfo{author}{\bibfnamefont{L.}~\bibnamefont{Neufcourt}},
  \bibinfo{author}{\bibfnamefont{Y.}~\bibnamefont{Cao}},
  \bibinfo{author}{\bibfnamefont{S.~A.} \bibnamefont{Giuliani}},
  \bibinfo{author}{\bibfnamefont{W.}~\bibnamefont{Nazarewicz}},
  \bibinfo{author}{\bibfnamefont{E.}~\bibnamefont{Olsen}}, \bibnamefont{and}
  \bibinfo{author}{\bibfnamefont{O.~B.} \bibnamefont{Tarasov}},
  \bibinfo{journal}{Phys. Rev. C} \textbf{\bibinfo{volume}{101}},
  \bibinfo{pages}{044307} (\bibinfo{year}{2020}).

\bibitem[{\citenamefont{Lovell et~al.}(2022)\citenamefont{Lovell, Mohan,
  Sprouse, and Mumpower}}]{Lovell:2022pkw}
\bibinfo{author}{\bibfnamefont{A.~E.} \bibnamefont{Lovell}},
  \bibinfo{author}{\bibfnamefont{A.~T.} \bibnamefont{Mohan}},
  \bibinfo{author}{\bibfnamefont{T.~M.} \bibnamefont{Sprouse}},
  \bibnamefont{and} \bibinfo{author}{\bibfnamefont{M.~R.}
  \bibnamefont{Mumpower}}, \bibinfo{journal}{Phys. Rev. C}
  \textbf{\bibinfo{volume}{106}}, \bibinfo{pages}{014305}
  (\bibinfo{year}{2022}).

\bibitem[{\citenamefont{Saito et~al.}(2024)\citenamefont{Saito, Dillmann,
  Kruecken, Mumpower, and Surman}}]{Saito:2023seh}
\bibinfo{author}{\bibfnamefont{Y.}~\bibnamefont{Saito}},
  \bibinfo{author}{\bibfnamefont{I.}~\bibnamefont{Dillmann}},
  \bibinfo{author}{\bibfnamefont{R.}~\bibnamefont{Kruecken}},
  \bibinfo{author}{\bibfnamefont{M.~R.} \bibnamefont{Mumpower}},
  \bibnamefont{and} \bibinfo{author}{\bibfnamefont{R.}~\bibnamefont{Surman}},
  \bibinfo{journal}{Phys. Rev. C} \textbf{\bibinfo{volume}{109}},
  \bibinfo{pages}{054301} (\bibinfo{year}{2024}).

\bibitem[{\citenamefont{Frame et~al.}(2018)\citenamefont{Frame, He, Ipsen, Lee,
  Lee, and Rrapaj}}]{Frame:2017fah}
\bibinfo{author}{\bibfnamefont{D.}~\bibnamefont{Frame}},
  \bibinfo{author}{\bibfnamefont{R.}~\bibnamefont{He}},
  \bibinfo{author}{\bibfnamefont{I.}~\bibnamefont{Ipsen}},
  \bibinfo{author}{\bibfnamefont{D.}~\bibnamefont{Lee}},
  \bibinfo{author}{\bibfnamefont{D.}~\bibnamefont{Lee}}, \bibnamefont{and}
  \bibinfo{author}{\bibfnamefont{E.}~\bibnamefont{Rrapaj}},
  \bibinfo{journal}{Phys. Rev. Lett.} \textbf{\bibinfo{volume}{121}},
  \bibinfo{pages}{032501} (\bibinfo{year}{2018}).

\bibitem[{\citenamefont{K\"onig et~al.}(2020)\citenamefont{K\"onig, Ekstr\"om,
  Hebeler, Lee, and Schwenk}}]{Konig:2019adq}
\bibinfo{author}{\bibfnamefont{S.}~\bibnamefont{K\"onig}},
  \bibinfo{author}{\bibfnamefont{A.}~\bibnamefont{Ekstr\"om}},
  \bibinfo{author}{\bibfnamefont{K.}~\bibnamefont{Hebeler}},
  \bibinfo{author}{\bibfnamefont{D.}~\bibnamefont{Lee}}, \bibnamefont{and}
  \bibinfo{author}{\bibfnamefont{A.}~\bibnamefont{Schwenk}},
  \bibinfo{journal}{Phys. Lett. B} \textbf{\bibinfo{volume}{810}},
  \bibinfo{pages}{135814} (\bibinfo{year}{2020}).

\bibitem[{\citenamefont{Drischler et~al.}(2021)\citenamefont{Drischler,
  Quinonez, Giuliani, Lovell, and Nunes}}]{Drischler:2021qoy}
\bibinfo{author}{\bibfnamefont{C.}~\bibnamefont{Drischler}},
  \bibinfo{author}{\bibfnamefont{M.}~\bibnamefont{Quinonez}},
  \bibinfo{author}{\bibfnamefont{P.~G.} \bibnamefont{Giuliani}},
  \bibinfo{author}{\bibfnamefont{A.~E.} \bibnamefont{Lovell}},
  \bibnamefont{and} \bibinfo{author}{\bibfnamefont{F.~M.} \bibnamefont{Nunes}},
  \bibinfo{journal}{Phys. Lett. B} \textbf{\bibinfo{volume}{823}},
  \bibinfo{pages}{136777} (\bibinfo{year}{2021}).

\bibitem[{\citenamefont{Bonilla et~al.}(2022)\citenamefont{Bonilla, Giuliani,
  Godbey, and Lee}}]{Bonilla:2022rph}
\bibinfo{author}{\bibfnamefont{E.}~\bibnamefont{Bonilla}},
  \bibinfo{author}{\bibfnamefont{P.}~\bibnamefont{Giuliani}},
  \bibinfo{author}{\bibfnamefont{K.}~\bibnamefont{Godbey}}, \bibnamefont{and}
  \bibinfo{author}{\bibfnamefont{D.}~\bibnamefont{Lee}},
  \bibinfo{journal}{Phys. Rev. C} \textbf{\bibinfo{volume}{106}},
  \bibinfo{pages}{054322} (\bibinfo{year}{2022}).

\bibitem[{\citenamefont{Melendez et~al.}(2022)\citenamefont{Melendez,
  Drischler, Furnstahl, Garcia, and Zhang}}]{Melendez:2022kid}
\bibinfo{author}{\bibfnamefont{J.~A.} \bibnamefont{Melendez}},
  \bibinfo{author}{\bibfnamefont{C.}~\bibnamefont{Drischler}},
  \bibinfo{author}{\bibfnamefont{R.~J.} \bibnamefont{Furnstahl}},
  \bibinfo{author}{\bibfnamefont{A.~J.} \bibnamefont{Garcia}},
  \bibnamefont{and} \bibinfo{author}{\bibfnamefont{X.}~\bibnamefont{Zhang}},
  \bibinfo{journal}{J. Phys. G} \textbf{\bibinfo{volume}{49}},
  \bibinfo{pages}{102001} (\bibinfo{year}{2022}).

\bibitem[{\citenamefont{Giuliani et~al.}(2023)\citenamefont{Giuliani, Godbey,
  Bonilla, Viens, and Piekarewicz}}]{Giuliani:2022yna}
\bibinfo{author}{\bibfnamefont{P.}~\bibnamefont{Giuliani}},
  \bibinfo{author}{\bibfnamefont{K.}~\bibnamefont{Godbey}},
  \bibinfo{author}{\bibfnamefont{E.}~\bibnamefont{Bonilla}},
  \bibinfo{author}{\bibfnamefont{F.}~\bibnamefont{Viens}}, \bibnamefont{and}
  \bibinfo{author}{\bibfnamefont{J.}~\bibnamefont{Piekarewicz}},
  \bibinfo{journal}{Front. Phys.} \textbf{\bibinfo{volume}{10}},
  \bibinfo{pages}{1054524} (\bibinfo{year}{2023}).

\bibitem[{\citenamefont{Anderson et~al.}(2022)\citenamefont{Anderson,
  O'Donnell, and Piekarewicz}}]{Anderson:2022jhq}
\bibinfo{author}{\bibfnamefont{A.~L.} \bibnamefont{Anderson}},
  \bibinfo{author}{\bibfnamefont{G.~L.} \bibnamefont{O'Donnell}},
  \bibnamefont{and}
  \bibinfo{author}{\bibfnamefont{J.}~\bibnamefont{Piekarewicz}},
  \bibinfo{journal}{Phys. Rev. C} \textbf{\bibinfo{volume}{106}},
  \bibinfo{pages}{L031302} (\bibinfo{year}{2022}).

\bibitem[{\citenamefont{Odell et~al.}(2024)\citenamefont{Odell, Giuliani,
  Beyer, Catacora-Rios, Chan, Bonilla, Furnstahl, Godbey, and
  Nunes}}]{Odell2024}
\bibinfo{author}{\bibfnamefont{D.}~\bibnamefont{Odell}},
  \bibinfo{author}{\bibfnamefont{P.}~\bibnamefont{Giuliani}},
  \bibinfo{author}{\bibfnamefont{K.}~\bibnamefont{Beyer}},
  \bibinfo{author}{\bibfnamefont{M.}~\bibnamefont{Catacora-Rios}},
  \bibinfo{author}{\bibfnamefont{M.~Y.-H.} \bibnamefont{Chan}},
  \bibinfo{author}{\bibfnamefont{E.}~\bibnamefont{Bonilla}},
  \bibinfo{author}{\bibfnamefont{R.~J.} \bibnamefont{Furnstahl}},
  \bibinfo{author}{\bibfnamefont{K.}~\bibnamefont{Godbey}}, \bibnamefont{and}
  \bibinfo{author}{\bibfnamefont{F.~M.} \bibnamefont{Nunes}},
  \bibinfo{journal}{Phys. Rev. C} \textbf{\bibinfo{volume}{109}},
  \bibinfo{pages}{044612} (\bibinfo{year}{2024}).

\bibitem[{\citenamefont{Godbey et~al.}()\citenamefont{Godbey, Giuliani,
  Bonilla, Flynn, Odell, Beyer, Lay, Figueroa, Garg, and Campbell}}]{drnuclear}
\bibinfo{author}{\bibfnamefont{K.}~\bibnamefont{Godbey}},
  \bibinfo{author}{\bibfnamefont{P.}~\bibnamefont{Giuliani}},
  \bibinfo{author}{\bibfnamefont{E.}~\bibnamefont{Bonilla}},
  \bibinfo{author}{\bibfnamefont{E.}~\bibnamefont{Flynn}},
  \bibinfo{author}{\bibfnamefont{D.}~\bibnamefont{Odell}},
  \bibinfo{author}{\bibfnamefont{K.}~\bibnamefont{Beyer}},
  \bibinfo{author}{\bibfnamefont{D.}~\bibnamefont{Lay}},
  \bibinfo{author}{\bibfnamefont{D.}~\bibnamefont{Figueroa}},
  \bibinfo{author}{\bibfnamefont{R.}~\bibnamefont{Garg}}, \bibnamefont{and}
  \bibinfo{author}{\bibfnamefont{M.}~\bibnamefont{Campbell}},
  \emph{\bibinfo{title}{Dimensionality reduction in nuclear physics}},
  \bibinfo{note}{\url{https://dr.ascsn.net/}}.

\bibitem[{\citenamefont{Brunton and Kutz}(2019)}]{Brunton_Kutz_2019}
\bibinfo{author}{\bibfnamefont{S.~L.} \bibnamefont{Brunton}} \bibnamefont{and}
  \bibinfo{author}{\bibfnamefont{J.~N.} \bibnamefont{Kutz}},
  \emph{\bibinfo{title}{Data-Driven Science and Engineering: Machine Learning,
  Dynamical Systems, and Control}} (\bibinfo{publisher}{Cambridge University
  Press}, \bibinfo{year}{2019}).

\bibitem[{\citenamefont{Benner et~al.}(2020)\citenamefont{Benner, Schilders,
  Grivet-Talocia, Quarteroni, Rozza, and Silveira}}]{Benner:2020}
\bibinfo{author}{\bibfnamefont{P.}~\bibnamefont{Benner}},
  \bibinfo{author}{\bibfnamefont{W.}~\bibnamefont{Schilders}},
  \bibinfo{author}{\bibfnamefont{S.}~\bibnamefont{Grivet-Talocia}},
  \bibinfo{author}{\bibfnamefont{A.}~\bibnamefont{Quarteroni}},
  \bibinfo{author}{\bibfnamefont{G.}~\bibnamefont{Rozza}}, \bibnamefont{and}
  \bibinfo{author}{\bibfnamefont{L.}~\bibnamefont{Silveira}},
  \emph{\bibinfo{title}{Model Order Reduction: Applications}}
  (\bibinfo{publisher}{De Gruyter}, \bibinfo{year}{2020}),
  vol.~\bibinfo{volume}{3}.

\bibitem[{\citenamefont{Quarteroni et~al.}(2015)\citenamefont{Quarteroni,
  Manzoni, and Negri}}]{Quarteroni:2015}
\bibinfo{author}{\bibfnamefont{A.}~\bibnamefont{Quarteroni}},
  \bibinfo{author}{\bibfnamefont{A.}~\bibnamefont{Manzoni}}, \bibnamefont{and}
  \bibinfo{author}{\bibfnamefont{F.}~\bibnamefont{Negri}},
  \emph{\bibinfo{title}{Reduced Basis Methods for Partial Differential
  Equations: An Introduction}}, vol.~\bibinfo{volume}{92}
  (\bibinfo{publisher}{Springer}, \bibinfo{year}{2015}).

\bibitem[{\citenamefont{Hesthaven et~al.}(2016)\citenamefont{Hesthaven, Rozza,
  and Stamm}}]{Heasthaven:2016}
\bibinfo{author}{\bibfnamefont{J.~S.} \bibnamefont{Hesthaven}},
  \bibinfo{author}{\bibfnamefont{G.}~\bibnamefont{Rozza}}, \bibnamefont{and}
  \bibinfo{author}{\bibfnamefont{B.}~\bibnamefont{Stamm}},
  \emph{\bibinfo{title}{Certified Reduced Basis Methods for Parametrized
  Partial Differential Equations}} (\bibinfo{publisher}{Springer},
  \bibinfo{address}{Switzerland}, \bibinfo{year}{2016}), vol.
  \bibinfo{volume}{590}.

\bibitem[{\citenamefont{Bohr and Mottelson}(1998)}]{BohrI:1998}
\bibinfo{author}{\bibfnamefont{A.}~\bibnamefont{Bohr}} \bibnamefont{and}
  \bibinfo{author}{\bibfnamefont{B.~R.} \bibnamefont{Mottelson}},
  \emph{\bibinfo{title}{Nuclear Structure}} (\bibinfo{publisher}{World
  Scientific Publishing Company, New Jersey}, \bibinfo{year}{1998}), vol.
  \bibinfo{volume}{I: Single-particle motion}.

\bibitem[{\citenamefont{Todd-Rutel and Piekarewicz}(2005)}]{Todd-Rutel:2005fa}
\bibinfo{author}{\bibfnamefont{B.~G.} \bibnamefont{Todd-Rutel}}
  \bibnamefont{and}
  \bibinfo{author}{\bibfnamefont{J.}~\bibnamefont{Piekarewicz}},
  \bibinfo{journal}{Phys. Rev. Lett} \textbf{\bibinfo{volume}{95}},
  \bibinfo{pages}{122501} (\bibinfo{year}{2005}).

\bibitem[{\citenamefont{Fattoyev et~al.}(2010)\citenamefont{Fattoyev, Horowitz,
  Piekarewicz, and Shen}}]{Fattoyev:2010mx}
\bibinfo{author}{\bibfnamefont{F.~J.} \bibnamefont{Fattoyev}},
  \bibinfo{author}{\bibfnamefont{C.~J.} \bibnamefont{Horowitz}},
  \bibinfo{author}{\bibfnamefont{J.}~\bibnamefont{Piekarewicz}},
  \bibnamefont{and} \bibinfo{author}{\bibfnamefont{G.}~\bibnamefont{Shen}},
  \bibinfo{journal}{Phys. Rev.} \textbf{\bibinfo{volume}{C82}},
  \bibinfo{pages}{055803} (\bibinfo{year}{2010}).

\bibitem[{\citenamefont{Chen and Piekarewicz}(2015)}]{Chen:2014mza}
\bibinfo{author}{\bibfnamefont{W.-C.} \bibnamefont{Chen}} \bibnamefont{and}
  \bibinfo{author}{\bibfnamefont{J.}~\bibnamefont{Piekarewicz}},
  \bibinfo{journal}{Phys. Lett.} \textbf{\bibinfo{volume}{B748}},
  \bibinfo{pages}{284} (\bibinfo{year}{2015}).

\bibitem[{\citenamefont{Hohenberg and Kohn}(1964)}]{Hohenberg:1964zz}
\bibinfo{author}{\bibfnamefont{P.}~\bibnamefont{Hohenberg}} \bibnamefont{and}
  \bibinfo{author}{\bibfnamefont{W.}~\bibnamefont{Kohn}},
  \bibinfo{journal}{Phys. Rev.} \textbf{\bibinfo{volume}{136}},
  \bibinfo{pages}{B864} (\bibinfo{year}{1964}).

\bibitem[{\citenamefont{Kohn and Sham}(1965)}]{Kohn:1965}
\bibinfo{author}{\bibfnamefont{W.}~\bibnamefont{Kohn}} \bibnamefont{and}
  \bibinfo{author}{\bibfnamefont{L.~J.} \bibnamefont{Sham}},
  \bibinfo{journal}{Phys. Rev.} \textbf{\bibinfo{volume}{140}},
  \bibinfo{pages}{A1133} (\bibinfo{year}{1965}).

\bibitem[{\citenamefont{Dirac}(1982)}]{Dirac:1982}
\bibinfo{author}{\bibfnamefont{P.}~\bibnamefont{Dirac}},
  \emph{\bibinfo{title}{The Principles of Quantum Mechanics}}
  (\bibinfo{publisher}{Clarendon Press, Oxford}, \bibinfo{year}{1982}).

\bibitem[{\citenamefont{Horowitz and Serot}(1981)}]{Horowitz:1981xw}
\bibinfo{author}{\bibfnamefont{C.~J.} \bibnamefont{Horowitz}} \bibnamefont{and}
  \bibinfo{author}{\bibfnamefont{B.~D.} \bibnamefont{Serot}},
  \bibinfo{journal}{Nucl. Phys.} \textbf{\bibinfo{volume}{A368}},
  \bibinfo{pages}{503} (\bibinfo{year}{1981}).

\bibitem[{\citenamefont{Serot and Walecka}(1986)}]{Serot:1984ey}
\bibinfo{author}{\bibfnamefont{B.~D.} \bibnamefont{Serot}} \bibnamefont{and}
  \bibinfo{author}{\bibfnamefont{J.~D.} \bibnamefont{Walecka}},
  \bibinfo{journal}{Adv. Nucl. Phys.} \textbf{\bibinfo{volume}{16}},
  \bibinfo{pages}{1} (\bibinfo{year}{1986}).

\bibitem[{\citenamefont{Sakurai}(1967)}]{Sakurai:1967}
\bibinfo{author}{\bibfnamefont{J.~J.} \bibnamefont{Sakurai}},
  \emph{\bibinfo{title}{Advanced Quantum Mechanics}}
  (\bibinfo{publisher}{Pearson Education}, \bibinfo{year}{1967}).

\end{thebibliography}

\end{document}